\documentclass[aps,prd,superscriptaddress,twocolumn,showpacs,floatfix]{revtex4-1}
\usepackage{graphicx}
\usepackage{dcolumn}
\usepackage{bm}
\usepackage{amsmath}
\usepackage[separate-uncertainty=true,load=addn]{siunitx}
\DeclareSIUnit\PSI{psi}
\DeclareSIUnit\PSIA{psia}
\usepackage{hyperref}
\usepackage{xcolor}
\hypersetup{
    colorlinks,
    linkcolor={red!50!black},
    citecolor={blue!50!black},
    urlcolor={blue!80!black}
}

\begin{document}

\preprint{APS/123-QED}

\title{Dark Matter Search Results from the Complete Exposure of the PICO-60 C$_3$F$_8$ Bubble Chamber} 

\author{C.~Amole}
\affiliation{Department of Physics, Queen's University, Kingston, K7L 3N6, Canada}

\author{M.~Ardid}
\affiliation{Departament de F\'isica Aplicada, IGIC - Universitat Polit\`ecnica de Val\`encia, Gandia 46730 Spain}

\author{I.~J.~Arnquist}
\affiliation{Pacific Northwest National Laboratory, Richland, Washington 99354, USA}

\author{D.~M.~Asner}
 \altaffiliation[now at ]{Brookhaven National Laboratory}
\affiliation{Pacific Northwest National Laboratory, Richland, Washington 99354, USA}

\author{D.~Baxter}
\affiliation{
Department of Physics and Astronomy, Northwestern University, Evanston, Illinois 60208, USA}
\affiliation{Enrico Fermi Institute, KICP, and Department of Physics, 
University of Chicago, Chicago, Illinois 60637, USA}

\author{E.~Behnke}
\affiliation{Department of Physics, Indiana University South Bend, South Bend, Indiana 46634, USA}

\author{M.~Bressler}
\affiliation{Department of Physics, Drexel University, Philadelphia, Pennsylvania 19104, USA}

\author{B.~Broerman}
\affiliation{Department of Physics, Queen's University, Kingston, K7L 3N6, Canada}

\author{G.~Cao}
\affiliation{Department of Physics, Queen's University, Kingston, K7L 3N6, Canada}

\author{C.~J.~Chen}
\affiliation{
Department of Physics and Astronomy, Northwestern University, Evanston, Illinois 60208, USA}

\author{U.~Chowdhury}
 \altaffiliation[now at ]{Canadian Nuclear Laboratories}
\affiliation{Department of Physics, Queen's University, Kingston, K7L 3N6, Canada}

\author{K.~Clark}
\affiliation{Department of Physics, Queen's University, Kingston, K7L 3N6, Canada}

\author{J.~I.~Collar}
\affiliation{Enrico Fermi Institute, KICP, and Department of Physics,
University of Chicago, Chicago, Illinois 60637, USA}

\author{P.~S.~Cooper}
\affiliation{Fermi National Accelerator Laboratory, Batavia, Illinois 60510, USA}

\author{C.~B.~Coutu}
\affiliation{Department of Physics, University of Alberta, Edmonton, T6G 2E1, Canada}

\author{C.~Cowles}
\affiliation{Pacific Northwest National Laboratory, Richland, Washington 99354, USA}

\author{M.~Crisler}
\affiliation{Fermi National Accelerator Laboratory, Batavia, Illinois 60510, USA}
\affiliation{Pacific Northwest National Laboratory, Richland, Washington 99354, USA}

\author{G.~Crowder}
\affiliation{Department of Physics, Queen's University, Kingston, K7L 3N6, Canada}

\author{N.~A.~Cruz-Venegas}
\affiliation{Department of Physics, University of Alberta, Edmonton, T6G 2E1, Canada}
\affiliation{Instituto de F\'isica, Universidad Nacional Aut\'onoma de M\'exico, M\'exico D.\:F. 01000, M\'exico}

\author{C.~E.~Dahl}
\affiliation{
Department of Physics and Astronomy, Northwestern University, Evanston, Illinois 60208, USA}
\affiliation{Fermi National Accelerator Laboratory, Batavia, Illinois 60510, USA}

\author{M.~Das}
\affiliation{Astroparticle Physics \& Cosmology Division, Saha Institute
of Nuclear Physics, HBNI, Kolkata, India}

\author{S.~Fallows}
 \email[Corresponding: ]{fallows@ualberta.ca}
\affiliation{Department of Physics, University of Alberta, Edmonton, T6G 2E1, Canada}

\author{J.~Farine}
\affiliation{Department of Physics, Laurentian University, Sudbury, P3E 2C6, Canada}

\author{I.~Felis}
\affiliation{Departament de F\'isica Aplicada, IGIC - Universitat Polit\`ecnica de Val\`encia, Gandia 46730 Spain}

\author{R.~Filgas}
\affiliation{Institute of Experimental and Applied Physics, Czech Technical University in Prague, Prague, Cz-12800, Czech Republic}

\author{F.~Girard}
\affiliation{Department of Physics, Laurentian University, Sudbury, P3E 2C6, Canada}
\affiliation{D\'epartement de Physique, Universit\'e de Montr\'eal, Montr\'eal, H3C 3J7, Canada}

\author{G.~Giroux}
\affiliation{Department of Physics, Queen's University, Kingston, K7L 3N6, Canada}

\author{J.~Hall}
\affiliation{SNOLAB, Lively, Ontario, P3Y 1N2, Canada}

\author{C.~Hardy}
\affiliation{Department of Physics, Queen's University, Kingston, K7L 3N6, Canada}

\author{O.~Harris}
\affiliation{Northeastern Illinois University, Chicago, Illinois 60625, USA}

\author{T.~Hillier}
\affiliation{Department of Physics, Laurentian University, Sudbury, P3E 2C6, Canada}

\author{E.~W.~Hoppe}
\affiliation{Pacific Northwest National Laboratory, Richland, Washington 99354, USA}

\author{C.~M.~Jackson}
\affiliation{Pacific Northwest National Laboratory, Richland, Washington 99354, USA}

\author{M.~Jin}
\affiliation{
Department of Physics and Astronomy, Northwestern University, Evanston, Illinois 60208, USA}

\author{L.~Klopfenstein}
\affiliation{Department of Physics, Indiana University South Bend, South Bend, Indiana 46634, USA}

\author{C.~B.~Krauss}
\affiliation{Department of Physics, University of Alberta, Edmonton, T6G 2E1, Canada}

\author{M.~Laurin}
\affiliation{D\'epartement de Physique, Universit\'e de Montr\'eal, Montr\'eal, H3C 3J7, Canada}

\author{I.~Lawson}
\affiliation{Department of Physics, Laurentian University, Sudbury, P3E 2C6, Canada}
\affiliation{SNOLAB, Lively, Ontario, P3Y 1N2, Canada}

\author{A.~Leblanc}
\affiliation{Department of Physics, Laurentian University, Sudbury, P3E 2C6, Canada}

\author{I.~Levine}
\affiliation{Department of Physics, Indiana University South Bend, South Bend, Indiana 46634, USA}

\author{C.~Licciardi}
\affiliation{Department of Physics, Laurentian University, Sudbury, P3E 2C6, Canada}

\author{W.~H.~Lippincott}
\affiliation{Fermi National Accelerator Laboratory, Batavia, Illinois 60510, USA}

\author{B.~Loer}
\affiliation{Pacific Northwest National Laboratory, Richland, Washington 99354, USA}

\author{F.~Mamedov}
\affiliation{Institute of Experimental and Applied Physics, Czech Technical University in Prague, Prague, Cz-12800, Czech Republic}

\author{P.~Mitra}
\affiliation{Department of Physics, University of Alberta, Edmonton, T6G 2E1, Canada}

\author{C.~Moore}
\affiliation{Department of Physics, Queen's University, Kingston, K7L 3N6, Canada}

\author{T.~Nania}
\affiliation{Department of Physics, Indiana University South Bend, South Bend, Indiana 46634, USA}

\author{R.~Neilson}
 \email[Corresponding: ]{neilson@drexel.edu}
\affiliation{Department of Physics, Drexel University, Philadelphia, Pennsylvania 19104, USA}

\author{A.~J.~Noble}
\affiliation{Department of Physics, Queen's University, Kingston, K7L 3N6, Canada}

\author{P.~Oedekerk}
\affiliation{Department of Physics, Indiana University South Bend, South Bend, Indiana 46634, USA}

\author{A.~Ortega}
\affiliation{Enrico Fermi Institute, KICP, and Department of Physics,
University of Chicago, Chicago, Illinois 60637, USA}

\author{M.-C.~Piro}
\affiliation{Department of Physics, University of Alberta, Edmonton, T6G 2E1, Canada}

\author{A.~Plante}
\affiliation{D\'epartement de Physique, Universit\'e de Montr\'eal, Montr\'eal, H3C 3J7, Canada}

\author{R.~Podviyanuk}
\affiliation{Department of Physics, Laurentian University, Sudbury, P3E 2C6, Canada}

\author{S.~Priya}
\affiliation{Materials Research Institute, Penn State, University Park, Pennsylvania 16802, USA}

\author{A.~E.~Robinson}
\affiliation{D\'epartement de Physique, Universit\'e de Montr\'eal, Montr\'eal, H3C 3J7, Canada}

\author{S.~Sahoo}
\affiliation{Astroparticle Physics \& Cosmology Division, Saha Institute
of Nuclear Physics, HBNI, Kolkata, India}

\author{O.~Scallon}
\affiliation{Department of Physics, Laurentian University, Sudbury, P3E 2C6, Canada}

\author{S.~Seth}
\affiliation{Astroparticle Physics \& Cosmology Division, Saha Institute
of Nuclear Physics, HBNI, Kolkata, India}

\author{A.~Sonnenschein}
\affiliation{Fermi National Accelerator Laboratory, Batavia, Illinois 60510, USA}

\author{N.~Starinski}
\affiliation{D\'epartement de Physique, Universit\'e de Montr\'eal, Montr\'eal, H3C 3J7, Canada}

\author{I.~\v{S}tekl}
\affiliation{Institute of Experimental and Applied Physics, Czech Technical University in Prague, Prague, Cz-12800, Czech Republic}

\author{T.~Sullivan}
\affiliation{Department of Physics, Queen's University, Kingston, K7L 3N6, Canada}

\author{F.~Tardif}
\affiliation{D\'epartement de Physique, Universit\'e de Montr\'eal, Montr\'eal, H3C 3J7, Canada}

\author{E.~V\'azquez-J\'auregui}
\affiliation{Instituto de F\'isica, Universidad Nacional Aut\'onoma de M\'exico, M\'exico D.\:F. 01000, M\'exico}
\affiliation{Department of Physics, Laurentian University, Sudbury, P3E 2C6, Canada}

\author{N.~Walkowski}
\affiliation{Department of Physics, Indiana University South Bend, South Bend, Indiana 46634, USA}

\author{E.~Weima}
\affiliation{Department of Physics, Laurentian University, Sudbury, P3E 2C6, Canada}

\author{U.~Wichoski}
\affiliation{Department of Physics, Laurentian University, Sudbury, P3E 2C6, Canada}

\author{K.~Wierman}
\affiliation{Pacific Northwest National Laboratory, Richland, Washington 99354, USA}

\author{Y.~Yan}
\affiliation{Materials Research Institute, Penn State, University Park, Pennsylvania 16802, USA}

\author{V.~Zacek}
\affiliation{D\'epartement de Physique, Universit\'e de Montr\'eal, Montr\'eal, H3C 3J7, Canada}

\author{J.~Zhang}
\altaffiliation[now at ]{Argonne National Laboratory}
\affiliation{
Department of Physics and Astronomy, Northwestern University, Evanston, Illinois 60208, USA}

\collaboration{PICO Collaboration}
\noaffiliation

\date{\today}

\begin{abstract}

Final results are reported from operation of the PICO-60 C$_3$F$_8$ dark matter detector, a bubble chamber filled with \SI{52}{\kg} of C$_3$F$_8$ located in the SNOLAB underground laboratory. The chamber was operated at thermodynamic thresholds as low as \SI{1.2}{\keV} without loss of stability. 
A new blind 1404-kg-day exposure at \SI{2.45}{\keV} threshold was acquired with approximately the same expected total background rate as the previous 1167-kg-day exposure at \SI{3.3}{\keV}. This increased exposure is enabled in part by a new optical tracking analysis to better identify events near detector walls, permitting a larger fiducial volume. 
These results set the most stringent direct-detection constraint to date on the WIMP-proton spin-dependent cross section at $2.5 \times 10^{-41}$\,cm$^2$ for a \SI{25}{\GeV} WIMP, and improve on previous PICO results for 3--\SI{5}{\GeV} WIMPs by an order of magnitude. 

\end{abstract}


\maketitle

\section{Introduction}


Identifying the particle nature of the cosmological dark matter is a central challenge in modern physics \cite{PDG,dmevidence,Jungman,Snowmass,wimpdetection}. Experiments attempting to directly detect Weakly Interacting Massive Particles (WIMPs) in the laboratory must be sensitive to the very small recoil energies (\SIrange[range-phrase=--,range-units=single]{1}{100}{\keV}) that WIMPs would deposit through elastic scattering on detector target nuclei of comparable mass. These detectors are designed to acquire large background-free exposures by using increasingly massive targets while minimizing all sources of backgrounds to a WIMP signal. The coupling between WIMPs and standard model particles is typically characterized in terms of spin-independent (SI) and spin-dependent (SD) cross sections. As the underlying mechanism for this interaction is unknown, a thorough WIMP-search program must probe both SI and SD couplings.

The superheated liquid detector technology used by the PICO collaboration 
affords excellent intrinsic rejection of electron recoils from gamma and beta particles. Alpha decays of U/Th daughter nuclei can be acoustically discriminated against using piezoelectric sensors mounted on the detector surface. Three dimensional optical event reconstruction allows for topological event selection, rejecting multiply scattering neutron events. Materials screening and optimized detector design minimize the sources of single-scatter neutron background, with the goal of acquiring a background-free WIMP-search exposure. 

The first blind exposure of the PICO-60 C$_3$F$_8$ detector \cite{30l_16_PRL} achieved this goal, acquiring a 1167-kg-day exposure at a thermodynamic threshold of \SI{3.3}{\keV} with zero single-scatter nuclear recoil candidates in the signal region after unblinding. 
Three multi-bubble events were observed during that exposure, while $0.25\pm 0.09$ single- and $0.96\pm 0.34$ multiple-scatter neutron events were predicted by simulation (Sec.~\ref{sec:bg}). This observation indicated that the detector was effectively neutron-limited, unable to attain significant additional WIMP sensitivity simply by acquiring longer exposures.

Following post-run calibrations, an attempt was made to explore the limits of detector stability at higher C$_3$F$_8$ temperatures and lower pressures, reducing the bubble nucleation threshold calculated using Equation 2 of Ref.~\cite{30l_13}. These thermodynamic changes were also expected to increase sensitivity to the environmental gamma background (Sec.~\ref{sec:bg}). The C$_3$F$_8$ temperature was increased from \SI{13.9(1)}{\celsius} to \SI{15.9(1)}{\celsius} and the superheated pressure was progressively reduced from \SI{30.2(3)}{\PSIA} to \SI{21.7(3)}{\PSIA}, effecting a reduction in the nucleation threshold (Sec.~\ref{sec:threshold}) from \SI{3.29(09)}{\keV} to \SI{1.81(08)}{\keV}. The detector continued to operate stably, maintaining a live-time fraction over 75\% during these periods, despite the higher rate of fiducial single-bubble events, as expected with increased sensitivity to the electron-recoil background.

In response, a second blind exposure was acquired between April and June 2017 at a threshold of \SI{2.45}{\keV}, for which the overall background rate was expected to be dominated by the same neutron background rates as at \SI{3.29}{\keV}. 
Here we report the results of that efficiency-corrected dark matter exposure of 1404\,kg-days.

Just prior to decommissioning, the temperature was raised to \SI{19.9(1)}{\celsius}, enabling thresholds as low as \SI{1.20(08)}{\keV} to be reached. As expected, the event rate was then dominated by events consistent with electron recoils, but operations remained stable. The higher event rate led to a reduced live-time fraction near 40\% at this lowest threshold. These operating conditions are summarized in Table~\ref{table:opcond}.

\section{Experimental Method}

\begin{figure}
\includegraphics[width=210 pt,trim=0 0 0 0,clip=true]{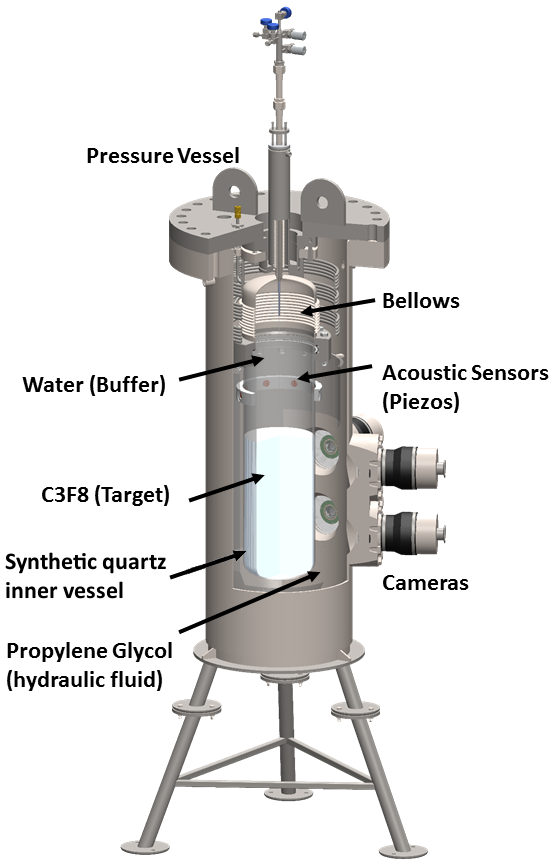}
\caption{\label{fig:pico60} The PICO-60 detector as configured for its operation with C$_3$F$_8$. The full volume of target fluid is stereoscopically imaged by two columns of two cameras each.} 
\end{figure}

The PICO-60 apparatus (Fig.~\ref{fig:pico60}) was configured as described in detail in~\cite{30l_13}, with the following changes implemented in 2016 for \cite{30l_16_PRL}. 
Rather than CF$_3$I, the bubble chamber was filled with \SI{52.2(5)}{\kg} of C$_3$F$_8$, 
as first reported in \cite{30l_16_PRL}. As the superheated operating temperatures for C$_3$F$_8$ are lower than those in CF$_3$I, a new chiller system was used to hold the temperature of the surrounding water tank \cite{30l_13,30l_16_PRL} uniform to approximately \SI{0.1}{\celsius}. 
The acoustic transducers, formerly coupled to the inner vessel with epoxy, were changed to a spring-loaded coupling. 
The chamber's expansion cycle from the stable, compressed state to the superheated state was identical to the previous run \cite{30l_16_PRL}, with only occasional minor alterations to the maximum cycle period and target pressure. These alterations were a response to temporarily elevated trigger rates observed after a temperature change, when thermal expansion or contraction of the C$_3$F$_8$ caused the position of its interface with the buffer water to shift, and visible water droplets to become localized sources of elevated wall nucleation rates. Relatively rapid cycling of the hydraulic system to an intermediate pressure over a period of approximately one hour was typically observed to return the chamber to stability.
These periods contain only diagnostic information and are neither blinded nor included in the present exposure. 

As in \cite{30l_16_PRL}, in order to image the entire C$_3$F$_8$ volume, double the volume used in PICO-60 CF$_3$I \cite{30l_13}, it was necessary to install an upper row of cameras, resulting in a stereoscopic view by each of two vertical pairs. As part of this expansion in scale, the data acquisition hardware and software running the cameras and issuing the primary event trigger was restructured and modularized prior to \cite{30l_16_PRL}. Each column of cameras was controlled by a separate server continuously acquiring images at \SI{340}{\Hz} for this result, improving time resolution compared to the \SI{200}{\Hz} used for the previous exposure \cite{30l_16_PRL}. Each camera filled a ring buffer with incoming images while its control software monitored for the appearance of bubbles by continuously calculating the difference-based spatial temporal entropy image \cite{DSTEI-Jing} $S_I=-\sum_i P_i\log_2 P_i$, where $P_i$ is the fraction of pixels populating intensity bin $i$ of the difference-map histograms generated from consecutive frames. 
These camera servers communicated operational state changes and trigger conditions to the primary data acquisition server managing event-level operation of the chamber. The cameras were sent a single digital pulse train to synchronize their exposure timing. This signal was also used to drive the pulse timing of the LEDs illuminating the chamber's inner volume.

 \begin{table*}[ht] 
 \begin{center} 
 \begin{tabular}{|c|c|c|c|c|} \hline
T ($^{\circ}$C) & P (psia) & Seitz threshold, $E_T$ (keV) & Livetime (d) & Exposure (kg-d) \\\hline  
  19.9 & 25.5 & $1.20 \pm 0.1(\mathrm{exp}) \pm 0.1(\mathrm{th})$&0.21 & 8.2 \\ 
  19.9 & 34.3 & $1.58 \pm 0.1(\mathrm{exp}) \pm 0.1(\mathrm{th})$&1.29 & 50.3 \\ 
  15.9 & 21.7 & $1.81 \pm 0.1(\mathrm{exp}) \pm 0.2(\mathrm{th})$&7.04 & 311 \\\hline\hline 
  15.9 & 30.5 & $2.45 \pm 0.1(\mathrm{exp}) \pm 0.2(\mathrm{th})$&29.95 & 1404 \\\hline 
  13.9 & 30.2 & $3.29 \pm 0.1(\mathrm{exp}) \pm 0.2(\mathrm{th})$&29.96 & 1167 \cite{30l_16_PRL}\\ 
\hline
\end{tabular}
 \caption{\label{table:opcond}
Details of the four new operating conditions and their associated exposures, as well as the original set of conditions used in \cite{30l_16_PRL}. The two blind exposures are grouped in the lower rows. The experimental uncertainty on the threshold comes from uncertainties on the temperature (\SI{0.1}{\celsius}) and pressure (0.3\,psi), while the theoretical uncertainty comes from the thermodynamic properties of C$_3$F$_8$ including the surface tension, and dominated by uncertainty in the Tolman length \cite{McLure}.}
 \end{center}
\end{table*}

The detector was primarily operated at four new sets of thermodynamic conditions, summarized in Table~\ref{table:opcond}. For the \SI{2.45(09)}{\keV} threshold, a second blind analysis \cite{30l_16_PRL} was undertaken by 
acquiring a new category of background data with masked acoustics. These acoustic signals allow discrimination between alpha decays and nuclear or electron recoil events with the Acoustic Parameter (AP) analysis variable, optimized to cleanly separate these distributions as in \cite{30l_16_PRL}. Source calibrations and pre-physics background data were used to finalize cuts and efficiencies for bulk single recoil event candidates in an unbiased way.
Unlike the previous blind analysis \cite{30l_16_PRL}, no supplemental neural network was used here to discriminate between alphas and nuclear recoils, though more advanced versions of this machine learning approach are being developed for future PICO detectors \cite{PICO_ML}.
After this analysis was frozen, acoustic information for the physics dataset was processed and the acceptance region was examined.

For the three lowest thresholds (1.20, 1.58, \SI{1.81}{\keV}), acoustic information was never blinded, and a full analysis not performed, as these datasets were always expected to contain many gamma-induced recoils indistinguishable from nuclear recoils by their acoustic signals. Furthermore, these lowest thresholds are not supported by comprehensive nuclear recoil calibrations in C$_3$F$_8$ as introduced for the thresholds of the blind exposures in Sec.~\ref{sec:threshold}.
These datasets thus act primarily
as a confirmation of the ability to operate a bubble chamber stably at very low thresholds, maintaining the superheated state for periods on the order of minutes, and are not included in the WIMP-search analysis.

\section{Bubble Nucleation Threshold}
\label{sec:threshold}

\begin{figure*}
\includegraphics[width=\textwidth,trim=0 0 0 0,clip=true]{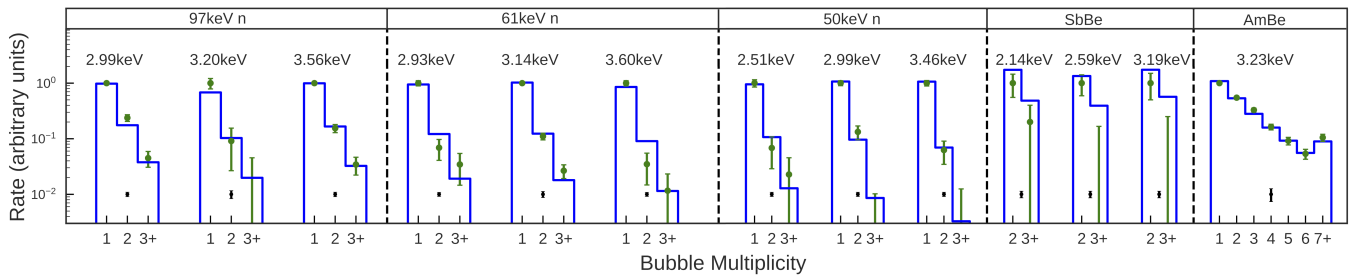}
\caption{\label{fig:multfit} 
Green points show the observed rates of single, double, and triple-or-more bubbles for the calibration sources at the listed thermodynamic thresholds. Green error bars indicate statistical uncertainties, and the black error bars at the bottom show the systematic uncertainty on the neutron flux. The blue histograms show the predicted rates from the simulation given the best-fit efficiency model derived from all calibration data. Each dataset is normalized to the observed rate of single bubbles, or double bubbles for SbBe due to gamma background.  The normalization of the simulation is constrained by the systematic neutron flux uncertainties shown.}
\end{figure*}

The efficiency with which nuclear recoils nucleate bubbles is measured with a suite of neutron calibration experiments, to which fluorine and carbon efficiency curves at each threshold are fit to monotonically increasing, piecewise linear functions. Well-defined resonances in the $^{51}$V(p,n)$^{51}$Cr reaction are used to produce monoenergetic 50, 61, and \SI{97}{\keV} neutrons directed at a $\sim$30-ml C$_3$F$_8$ bubble chamber at the Tandem Van de Graaff facility at the Universit\'e de Montr\'eal. An SbBe neutron source is also deployed adjacent to the $\sim$30-ml bubble chamber, and an AmBe neutron source adjacent to the PICO-2L chamber~\cite{2l_13}. The initial C$_3$F$_8$ calibration presented in Ref.~\cite{2l_13} and used for the first PICO-60 C$_3$F$_8$ result~\cite{30l_16_PRL} is refined in this analysis with additional calibration data. Datasets have been extended for 61 and \SI{97}{\keV} neutron beams and the \SI{50}{\keV} neutron beam dataset is entirely new, as is the SbBe source, a gamma-induced neutron source that primarily produces monoenergetic \SI{24}{\keV} neutrons. Calibrations were performed at a variety of thermodynamic thresholds, with selected results shown in Figure \ref{fig:multfit}, along with the prediction for the best-fit efficiency model.

\begin{figure}
\includegraphics[width=250 pt,trim=0 2cm 0 2cm,clip=true]{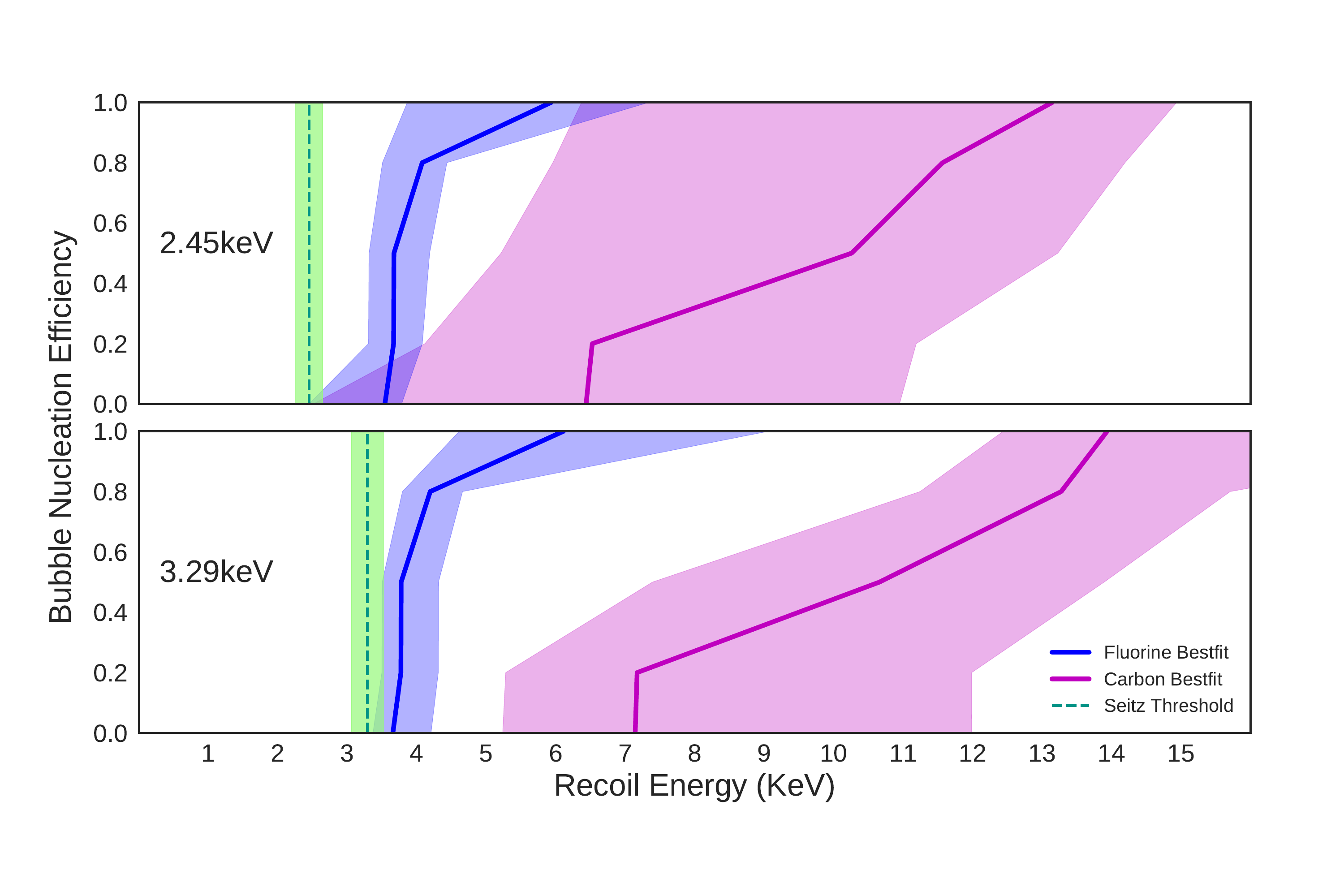}
\caption{\label{fig:efficiency} Best-fit fluorine (blue) and carbon (magenta) efficiency curves for \SI{2.45}{\keV} and \SI{3.29}{\keV} data are shown as solid lines. The shaded regions show the band enveloping all efficiency curves fitted within 1$\sigma$. The green dashed lines show the calculated Seitz threshold, with theoretical uncertainties from Table~\ref{table:opcond}.}
\end{figure}

Each of the neutron calibration experiments is simulated in MCNP~\cite{MCNP} or Geant4~\cite{GEANT4}, using differential cross sections for elastic scattering on fluorine from Ref.~\cite{Robinson}. The calibration data is fit using the \texttt{emcee}~\cite{emcee} Markov Chain Monte Carlo (MCMC) python code package. The output of the fitting is a distribution of sets of four efficiency curves (fluorine and carbon curves at each of the 2.45 and 3.29~keV thresholds) with associated likelihoods (Fig.~\ref{fig:efficiency}). The addition of the new lower-energy neutron datasets supports tighter constraints on the low-energy part of the efficiency curves than previously reported, resulting in increased sensitivity to low-mass WIMPs. A detailed paper on the calibration of the bubble nucleation efficiency is in preparation by the collaboration \cite{PICO_NR}.

\section{Data Analysis}

The new datasets were processed as in \cite{30l_16_PRL} with independently redefined cuts for each set of operating conditions, and with several improvements and additions.

For this analysis a more sophisticated determination of the fiducial volume was deployed, using better bubble position reconstruction and a tag for wall-originating events based on bubble motion.
The position reconstruction algorithm was modified to have finer, sub-pixel resolution, and to monitor and correct for small shifts in the overall image position on a camera's sensor over time. Each camera's contribution to the reconstruction was individually weighted by the relative quality of the bubble image obtained. Image quality was evaluated as a function of the distance between the bubble's image and the center of the camera's sensor, and included corrections for lighting quality that changed as several LEDs failed during operation.

A new tracking algorithm supplemented information about the bubble position at the time of first appearance with its position across up to nine successive frames, over a total period of \SI{30}{\milli\second}. Bubbles nucleated on the walls were typically observed to follow tracks angled 10$^{\circ}$ or more from vertical over this period and could be rejected. This tagging by zenith angle had 95.6\% acceptance of non-wall events in the annular region \SIrange[range-phrase=--,range-units=single]{3}{7}{\mm} from the wall where it was applied. 
Similarly, events near the surface (the top \SI{10}{\mm} of active fluid), where visibility is less favorable, were required to be detected by both cameras within \SI{30}{\milli\second} of each other, to limit uncertainties in position reconstruction. This cut had 100\% acceptance of non-surface events in the cylindrical near-surface region.

Together, these optimizations allow fiducial cut boundaries to be placed closer to the edge of the detector while still classifying zero surface- or wall-originating events as fiducial. The fiducial mass is thus increased relative to \cite{30l_16_PRL} from ($45.7\pm0.5$)\,\si{\kg} to ($48.9\pm 0.8$)\,\si{\kg}. 
Together with a higher singles selection efficiency than \cite{30l_16_PRL} due to a slightly wider AP cut (Fig.~\ref{fig:APvsNN}) and lack of a neural network-based acoustic cut, this results in a WIMP-search exposure of \SI{1404}{\kg\day} for the second blind run of PICO-60 C$_3$F$_8$, as detailed in Table~\ref{table:runinfo}. 

\begin{figure}
\includegraphics[width=240 pt,trim=0 0 0 0,clip=true]{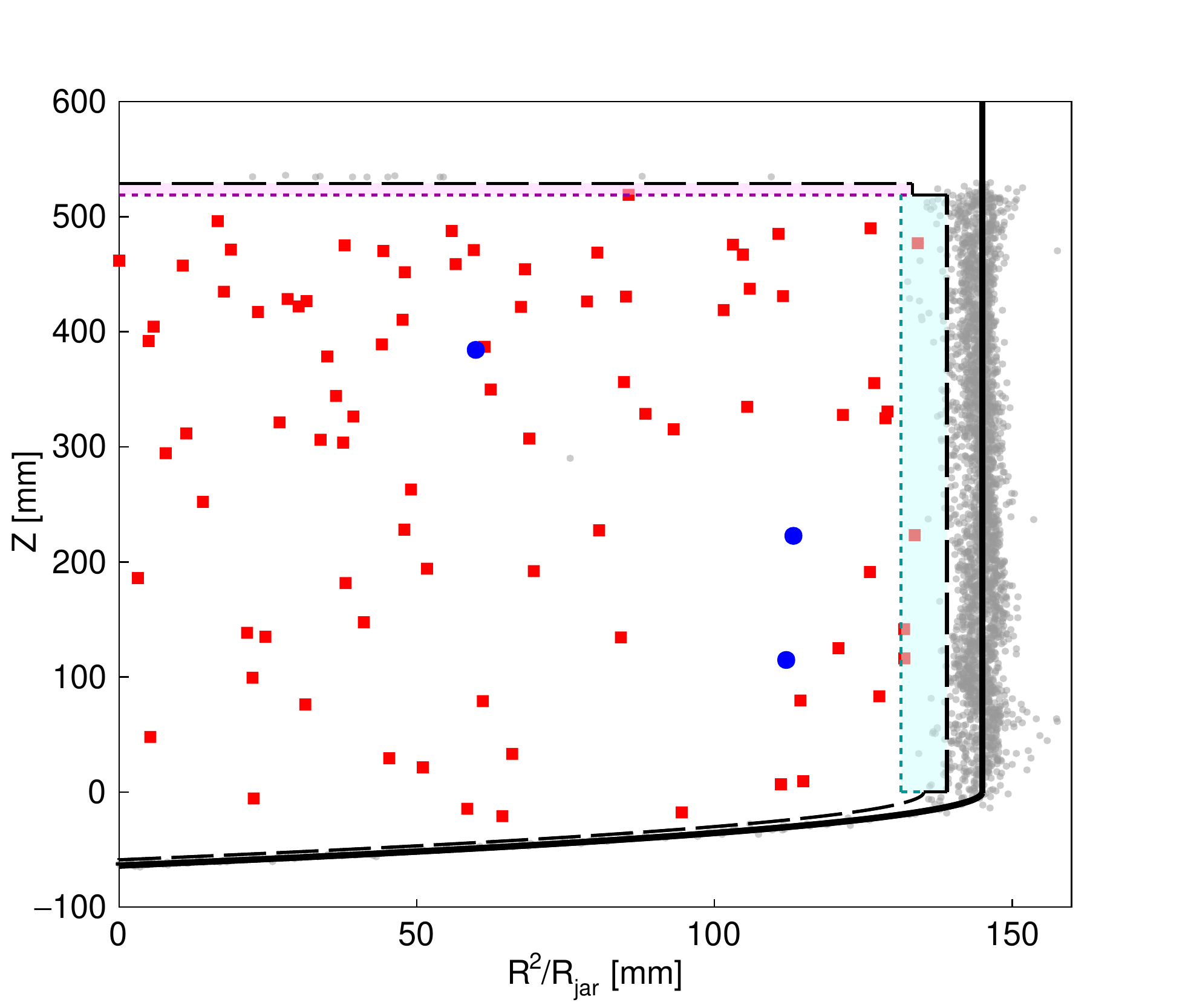}
\caption{\label{fig:XYZ} Spatial distribution of single-bubble events in the \SI{2.45}{\keV} WIMP-search data. Z is the reconstructed vertical position of the bubble, and R$^{2}$/R$_{\mathrm{jar}}$ is the distance from the center axis squared, normalized by the nominal jar radius (\SI{145}{\mm}). The outer edge of the fiducial cut is represented by the dashed black line, outside of which all events are excluded. Events reconstructed within the cyan annular region \SIrange[range-phrase=--,range-units=single]{3}{7}{\mm} from the wall were additionally required to satisfy a condition limiting their track's zenith angle, and events in the \SI{10}{\mm} near-surface magenta region were additionally required to have appeared in two cameras with a limited offset in frame index.
Red squares are the 87 single bulk bubbles passing all cuts prior to acoustic unblinding and grey dots are all rejected single-bubble events. Blue circles are the 3 candidate events passing the AP cut.}
\end{figure}

Time-dependent effects over the blind exposure were minimal. As in the past the rate of pressure rise during early stages of bubble growth, measured by a Dytran 2005V fast pressure transducer \cite{dytran}, was used to identify bubble multiplicity. The Dytran signal drifted slowly over time and was renormalized in the analysis. The magnitude of acoustic signals from bubble nucleation is strongly dependent on the temperature and pressure of the superheated liquid. Single-bubble events from $^{252}$Cf neutron and $^{133}$Ba gamma calibration data taken at each set of thermodynamic conditions were used to create separately normalized AP definitions for those conditions.

\section{Background estimates}
\label{sec:bg}

Backgrounds are estimated from a combination of Monte Carlo simulations, measured calibration event rates, and multi-bubble event rates during physics runs. 
The total expected background rate is the sum of the following contributions and is summarized in Table~\ref{table:bg_estimates}.

The majority of neutron scatter events induce more than one visible bubble in the detector, unlike single-bubble WIMP-scattering events. Multi-bubble events provide a definitive signature of neutron background and represent the most robust constraint on the rate of single-scatter neutron background events. Geant4 simulations of the detector geometry and composition predict 80\% of neutron events to have multiple bubbles, in agreement with $^{252}$Cf neutron calibration data, and with minimal dependence on the type and location of the neutron source. Since the detector configuration was unchanged between the first and second blind exposures (and since multiplicity was not blinded), the neutron rate is estimated from the overall rate of multi-bubble events from both exposures. Five multi-bubble events were observed, three in the first blind exposure~\cite{30l_16_PRL} and two in the second blind exposure, resulting is a neutron background expectation for the \SI{2.45} {keV} exposure of 0.8\,$\pm$\,0.4 events. The observed multi-bubble rate is modestly higher than predicted from the simulations, which estimate 0.96\,$\pm$\,0.34 and 1.43\,$\pm$\,0.49 multiples in the \SI{3.29} {\keV} and \SI{2.45} {\keV} exposures respectively, and 0.38\,$\pm$\,0.15 single-scatter background events in the \SI{2.45} {\keV} exposure. The discrepancy between the observed and predicted multi-bubble event rate is not significant at the 90\% C.L. The observed multi-bubble event rate is used to calculate a data-driven prediction of the single-bubble neutron background shown in Table~\ref{table:bg_estimates}.

Gamma calibration was performed at \SI{2.45}{\keV} with a \SI{0.1}{\milli\curie} $^{60}$Co source before and after the blinded run. Compared with a Geant4 simulation of the same detector geometry, this produces a measured nucleation efficiency of (2.89 $\pm$ 0.15)$\times 10^{-9}$ for gamma interaction events producing electron recoils above \SI{2.45}{\keV}. Combined with the rate from external gammas as simulated in MCNP, based on measurements from a NaI detector close to PICO-60 at SNOLAB
\cite{robinson_thesis,fustin_thesis}, we estimate a background of 
0.12 $\pm$ 0.02 gamma events in this 1404\,kg-day 
blind exposure.
More advanced models of the gamma response in superheated fluids are currently under development by the PICO collaboration \cite{PICO_ER}.

The rate of coherent elastic nuclear scattering of $^8$B solar neutrinos on C$_3$F$_8$ is non-negligible for thresholds below approximately \SI{5}{\keV}, so we calculate this contribution to the total background rate.
For the blind exposure acquired at \SI{2.45}{\keV}, this background is projected to contribute (0.10 $\pm$ 0.02) events. 

The measured fiducial single-bubble event rate during the second blind run of PICO-60 C$_3$F$_8$, (2.9 $\pm$ 0.3) events/live-day, can be extrapolated to a $^{222}$Rn rate under the assumption that each such event represents one of three alpha decays along the $^{222}$Rn to $^{210}$Po chain. Given the exposure of this dataset, this corresponds to an approximate $^{222}$Rn rate of \SI{2}{\micro\becquerel} in the detector, competitive with DEAP-3600~\cite{deap3600}.
In our \SI{2.45}{\keV} blind exposure, excellent separation of low-AP recoil events from radon chain alphas is maintained.
We therefore assume zero contribution to the total background rate from these events.

\begin{table}
\centering
\begin{tabular}{| c | c | c | c |}
\hline
Neutron & Gamma & CE$\nu$NS & Total \\\toprule\hline
(0.8 $\pm$ 0.4) & (0.12 $\pm$ 0.02) & (0.10 $\pm$ 0.02) & (1.0 $\pm$ 0.4)\\\hline
\end{tabular}
\caption{Summary of estimated single-bubble background contributions for the full 29.95 day livetime of the \SI{2.45}{\keV} blind run of PICO-60 C$_3$F$_8$. ``CE$\nu$NS'' indicates the contribution from coherent elastic neutrino-nucleus scattering.}
\label{table:bg_estimates}
\end{table}

\section{WIMP search Results}

\begin{figure}
\includegraphics[width=240 pt,trim=0.2cm 0.2cm 0.9cm 0.5cm,clip=true]{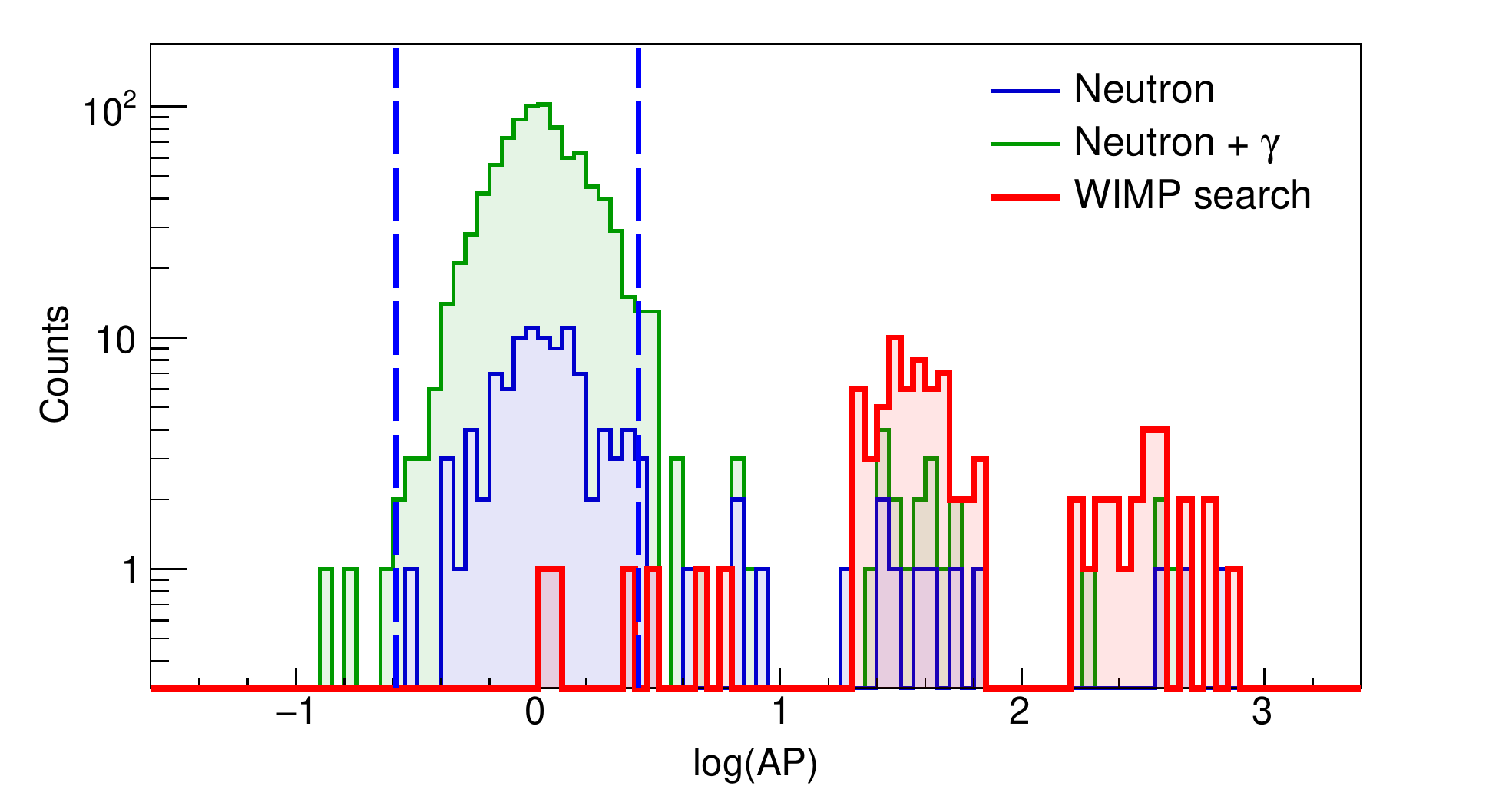}
\caption{\label{fig:APvsNN} AP distributions for $^{252}$Cf (blue) and $^{133}$Ba calibration data (combined with $^{252}$Cf, green) and WIMP-search data (red) at \SI{2.45}{\keV} threshold. The acceptance region for nuclear recoil candidates, defined before WIMP-search acoustic data unmasking using neutron and gamma calibration data to span \mbox{(--3$\sigma$,} \mbox{+2$\sigma$)} from the mean, is displayed with dashed blue lines, and reveals 3 candidate events in the WIMP-search data. Alphas from the $^{222}$Rn decay chain can be identified by their time signature and populate the two peaks in the WIMP-search data at high AP. Higher energy alphas from $^{214}$Po produce larger acoustic signals.}
\end{figure}

After the decision to unblind the \SI{2.45}{\keV} WIMP-search dataset, the acoustic signals were processed and are presented in Figure~\ref{fig:APvsNN} along with the AP distributions for neutron and gamma calibrations. Three nuclear recoil candidates are observed in the WIMP-search signal region, consistent with the background prediction from Table~\ref{table:bg_estimates} at the 90\% C.L. The total observation of three single-bubble and five multiple-bubble events over the combined exposures is consistent with the expected singles-to-multiples ratio of 1:4 for a neutron dominated background, albeit at somewhat higher rate than predicted by our simulations.

\begin{table*}
\begin{center}
\begin{tabular*}{\textwidth}{  l @{\extracolsep{\fill}} c c c c }
\hline 
\hline
\rule{0pt}{2.5ex}Dataset & Efficiency ($\%$) & Fiducial Mass (kg) & Exposure (kg-days) & Number of events \\
\hline
\rule{0pt}{2.5ex}Singles & 95.9 $\substack{+1.9\\-3.4}$ & 48.9 $\pm$ 0.8 & 1404 $\substack{+48\\-75}$ & 3 \\
\rule{0pt}{2.5ex}Multiples & 99.9 $\substack{+0.0\\-0.1}$ & 52.0 $\pm$ 0.1 & 1556 $\substack{+3\\-5}$ & 2 \\ \hline
\hline
\end{tabular*}
\caption{Summary of the final number of events and exposure determination for singles and multiples in the 29.95 live-day WIMP-search dataset of PICO-60 C$_{3}$F$_{8}$ at \SI{2.45}{\keV} thermodynamic threshold. The singles selection efficiency is substantially higher than that of \cite{30l_16_PRL} due to a slightly wider AP acceptance region and the omission of the supplemental neural network-based acoustic cut used in the prior analysis.}
\label{table:runinfo}
\end{center}
\end{table*}

A Profile Likelihood Ratio (PLR) test~\cite{plr} is used to set WIMP exclusion limits on the combination of 2.45 and 3.29~keV datasets. A test statistic is formed from the ratio of the likelihood for a specific WIMP cross-section to the maximum likelihood over all WIMP cross-sections. The background rate and WIMP detection efficiency in each dataset are treated as nuisance parameters, marginalized over by finding the conditional maximum likelihood for each specific WIMP cross-section. 

Given the apparent discrepancy between our predicted and observed neutron background, the background rates are unconstrained in the PLR, with flat likelihood functions for all non-negative values. In future PICO dark matter searches the neutron background rate may be constrained in the PLR by including the multi-bubble event rate, but to be conservative that has not been implemented in this analysis.

For the efficiencies, a likelihood surface is created as a function of WIMP detection efficiency at 2.45 and 3.29~keV. WIMP detection efficiencies, $\Phi$, in units of detected WIMPs per kg-day of exposure per picobarn of WIMP-nucleon scattering cross-section, are derived from the calibration MCMC output by integrating the efficiency curves over the nuclear recoil spectrum from an astrophysical WIMP flux for an array of potential WIMP masses.  The two-dimensional WIMP detection efficiency space is divided into bins and within each bin the maximum likelihood set of efficiency curves that fall within that bin is found. The likelihood surface thus created retains any covariance between the efficiency at the two thresholds from the neutron calibration. 

\begin{figure}
\includegraphics[width=280 pt,trim=0 0 0 0,clip=true]
{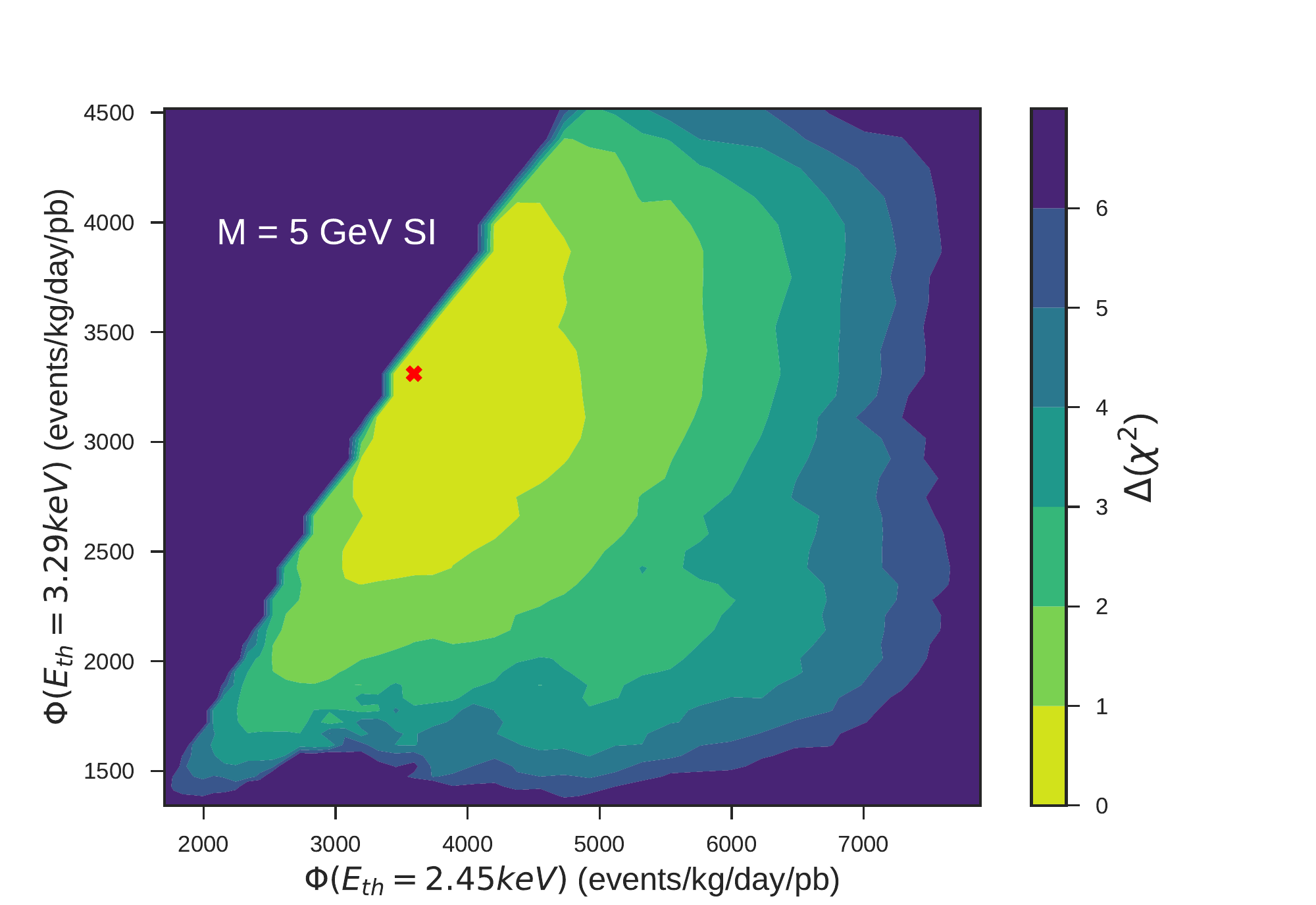}
\includegraphics[width=280 pt,trim=0 0 0 0,clip=true]
{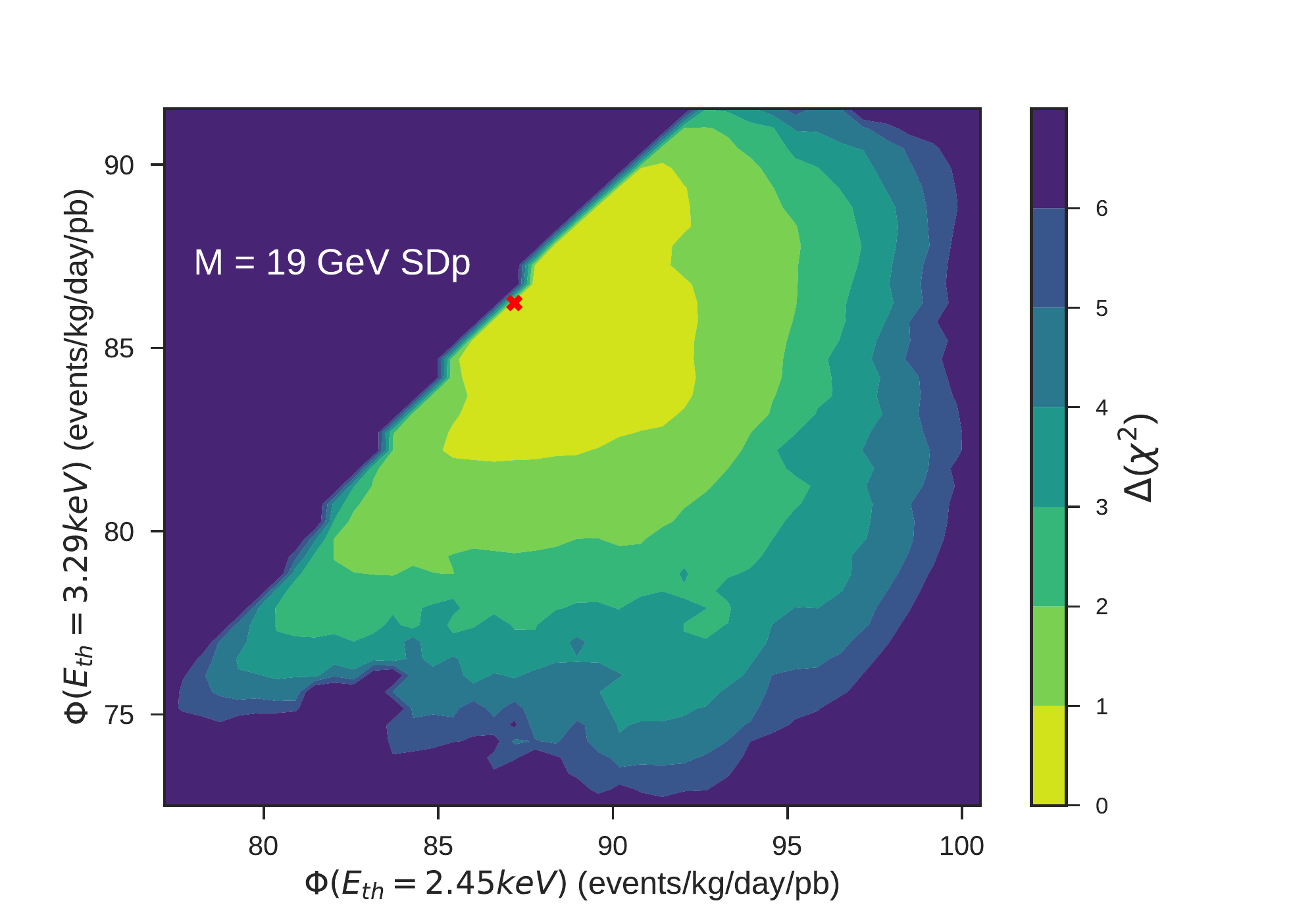}
\caption{\label{fig:contour} Contour plot of integrated efficiency $\Phi$ at \SI{2.45}{\keV} and \SI{3.29}{\keV} with red dot representing best-fit result. Contour layers have been color-coded to represent the difference in $\chi^2$ with respect to the minimum. Details in the outer boundary of the plot are subject to statistical fluctuations.}
\end{figure}

The standard halo parametrization~\cite{lewinandsmith} is used, with the following parameters: local dark matter density $\rho_D$=\SI{0.3}{\GeV}$c^{-2}$cm$^{-3}$, galactic escape velocity $v_{\rm{esc}}$ = 544 km/s, velocity of the earth with respect to the halo $v_{\rm Earth}$ = 232 km/s, and characteristic WIMP velocity with respect the halo $v_0$ = 220 km/s. The effective field theory treatment and nuclear form factors described in Refs.~\cite{spindependentcouplings,Anand,Gresham,Gluscevic} are used to determine sensitivity to both spin-dependent and spin-independent dark matter interactions. The $M$ response of Table 1 in Ref.~\cite{spindependentcouplings} is used for SI interactions, and the sum of the $\Sigma'$ and $\Sigma''$ terms from the same table is used for SD interactions. To implement these interactions and form factors, the publicly available \texttt{dmdd} code package~\cite{Gluscevic,Gluscevic2} is used. Figure~\ref{fig:contour} shows examples of the WIMP detection efficiency likelihood surfaces used for 5\,GeV WIMPs with SI coupling and 19\,GeV WIMPs with SD-proton coupling. The likelihood surfaces are marginalized over as nuisance parameters within the PLR, after being convolved with a two-dimensional Gaussian function reflecting experimental uncertainty in the PICO-60 thermodynamic thresholds. 

\begin{figure}
\includegraphics[width=240 pt]{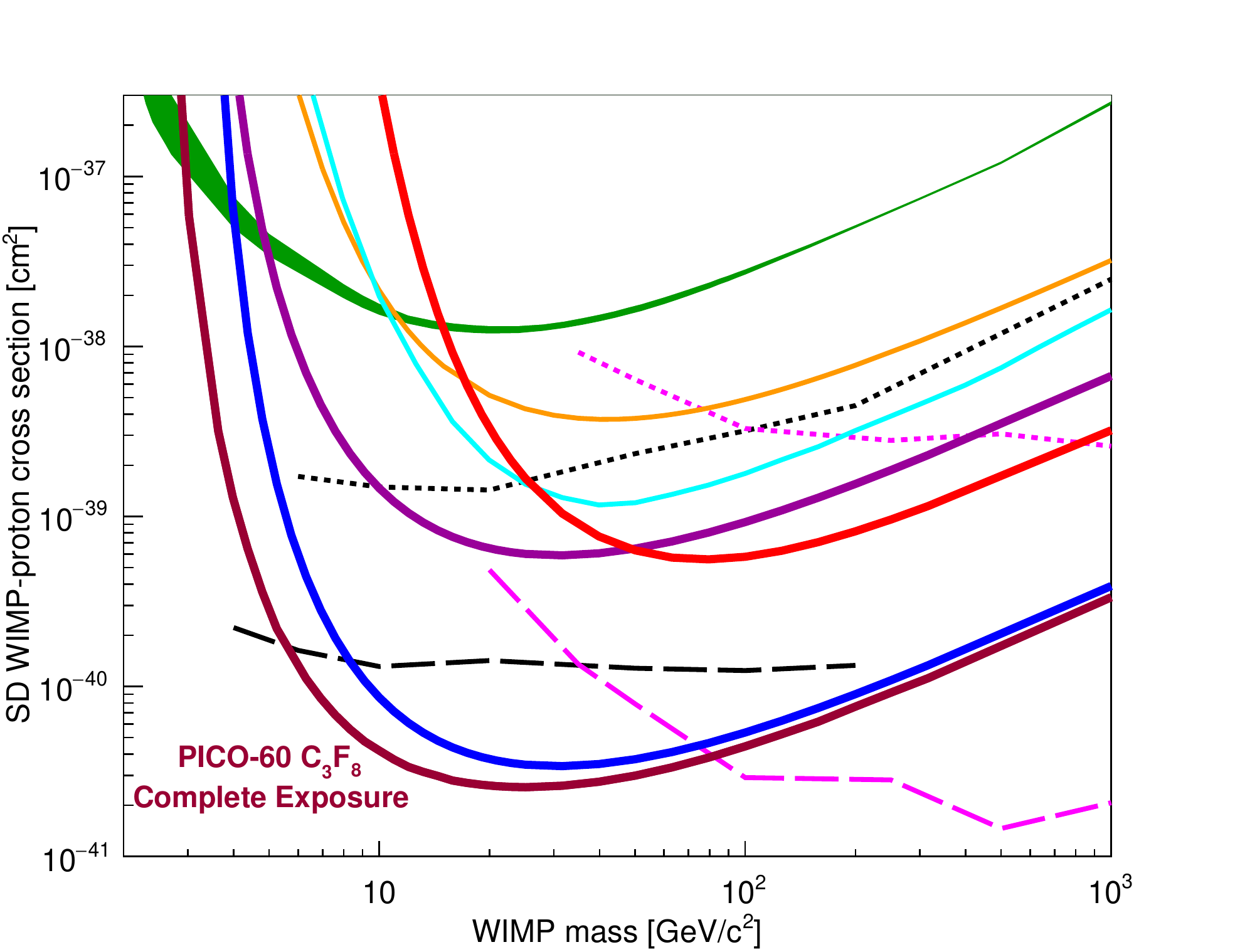}
\caption{\label{fig:SD} The  90\% C.L. limit on the SD WIMP-proton cross section from the profile likelihood analysis of the PICO-60 C$_3$F$_8$ combined blind exposure plotted in thick maroon, along with limits from the first blind exposure of PICO-60 C$_3$F$_8$ (thick blue) \cite{30l_16_PRL}, as well as limits from PICO-60 CF$_3$I (thick red)~\cite{30l_13}, PICO-2L (thick purple)~\cite{2l_15}, PICASSO (green band)~\cite{PICASSOFinallimit}, SIMPLE (orange)~\cite{SIMPLE}, PandaX-II (cyan)~\cite{PANDAX-II}, IceCube (dashed and dotted pink)~\cite{ICECUBElimit}, and SuperK (dashed and dotted black)~\cite{SKlimit,SKlimit2}. The indirect limits from IceCube and SuperK assume annihilation to $\tau$ leptons (dashed) and {\it b} quarks (dotted). 
Additional limits, not shown for clarity, are set by LUX~\cite{LUX_SD} and XENON1T~\cite{XENON1T} (comparable to PandaX-II) and by ANTARES~\cite{Ant1,Ant2} (comparable to IceCube).}
\end{figure}

\begin{figure}
\includegraphics[width=240 pt,trim=0 0 25 15,clip=true]{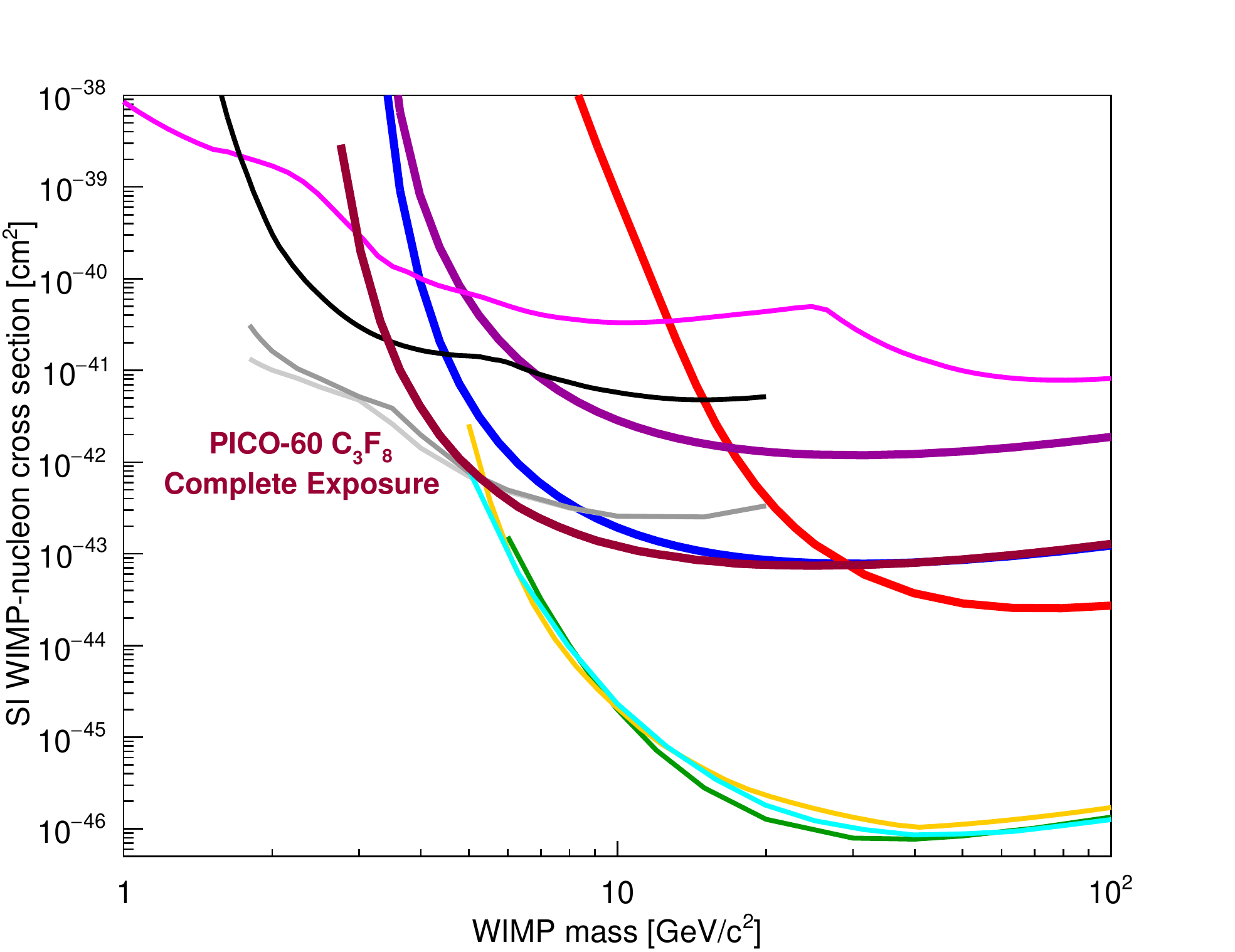}
\caption{\label{fig:SI} The 90\% C.L. limit on the SI WIMP-nucleon cross-section from the profile likelihood ratio analysis of the PICO-60 C$_3$F$_8$ combined blind exposure plotted in thick maroon, along with limits from the first blind exposure of PICO-60 C$_3$F$_8$ (thick blue) \cite{30l_16_PRL}, PICO-60 CF$_3$I (thick red)~\cite{30l_13}, PICO-2L (thick purple)~\cite{2l_15}, DarkSide-50 low-mass (gray)~\cite{DarkSide50LM}, LUX (yellow)~\cite{LUX2017}, PandaX-II (cyan)~\cite{PANDAX-II_SI}, XENON1T (green) \cite{XENON1T}, CRESST-II (magenta)~\cite{CRESST}, and CDMS-lite (black)~\cite{CDMSlite}. 
Additional limits, not shown for clarity, are set by PICASSO~\cite{PICASSOFinallimit}, XENON100~\cite{XENON100}, SuperCDMS~\cite{SuperCDMS}, CDMS-II~\cite{CDMSII}, and Edelweiss-III~\cite{Edelweiss}. 
}
\end{figure}

To develop a frequentist WIMP exclusion curve, toy datasets are generated at each point in a grid of WIMP masses and cross-sections. A grid point is then excluded if the observed PLR test statistic for that point is $>$90\% of toy dataset test statistics at that point. A conservative choice is made to generate the toy datasets with no background contribution, but the 90\% exclusion curve is subsequently confirmed to be valid over the range of background rates consistent with the data. The calculated exclusion curves at 90\% C.L. for spin-dependent WIMP-proton and spin-independent WIMP-nucleon elastic scattering cross-sections, as a function of WIMP mass, are shown in Figures~\ref{fig:SD} and~\ref{fig:SI}. The already world-leading limits in the spin-dependent WIMP-proton sector are improved, particularly for WIMP-masses in the \SIrange[range-phrase=--,range-units=single]{3}{5}{\GeV} range.

\section{Acknowledgements}

The PICO collaboration wishes to thank SNOLAB and its staff for support through underground space, logistical and technical services. SNOLAB operations are supported by the Canada Foundation for Innovation and the Province of Ontario Ministry of Research and Innovation, with underground access provided by Vale at the Creighton mine site. We wish to acknowledge the support of the Natural Sciences and Engineering Research Council of Canada (NSERC) and the Canada Foundation for Innovation (CFI) for funding. We acknowledge the support from the National Science Foundation (NSF) (Grant 0919526, 1506337, 1242637 and 1205987). We acknowledge that this work is supported by the U.S.\ Department of Energy (DOE) Office of Science, Office of High Energy Physics (under award DE-SC-0012161), by the DOE Office of Science Graduate Student Research (SCGSR) award,  by DGAPA-UNAM (PAPIIT No.\:IA100316 and IA100118) and Consejo Nacional de Ciencia y Tecnolog\'ia (CONACyT, M\'exico, Grant No.\:252167), by the Department of Atomic Energy (DAE), Government of India, under the Centre for AstroParticle Physics II project (CAPP-II) at the Saha Institute of Nuclear Physics (SINP), 
European Regional Development Fund-Project ``Engineering applications of microworld physics'' (No.\:CZ.02.1.01/0.0/0.0/16\_019/0000766), 
and 
the Spanish Ministerio de Ciencia, Innovaci\'on y Universidades (Red Consolider MultiDark, FPA2017$-$90566$-$REDC).
This work is partially supported by the Kavli Institute for Cosmological Physics at the University of Chicago through NSF grant 1125897 and 1806722, and an endowment from the Kavli Foundation and its founder Fred Kavli. We also wish to acknowledge the support from Fermi National Accelerator Laboratory under Contract No.\:DE-AC02-07CH11359, and from Pacific Northwest National Laboratory, which is operated by Battelle for the U.S.\ Department of Energy under Contract No.\:DE-AC05-76RL01830. We also thank Compute Canada (\url{www.computecanada.ca}) and the Centre for Advanced Computing, ACENET, Calcul Qu\'ebec, Compute Ontario and WestGrid for computational support.

\end{document}